\documentclass[aip,preprint]{revtex4-1}
\usepackage{amsmath}
\usepackage{graphicx}
\usepackage{bm}

\begin{document}

\title{Depletion interaction between two ellipsoids}

\author{Han Miao$^{a}$ ,Yao Li$^{b, a}$, Hongru Ma$^{c}$}

\affiliation{$^{a}$ Department of Physics, Shanghai Jiao Tong University, Shanghai
200240 China}

\affiliation{$^{b}$ Department of Physics, Tsinghua University, Beijing 100084,
China}

\affiliation{$^{c}$ School of Mechanical Engineering, and Key Laboratory of Artificial Structures and Quantum Control (Ministry of Education), Shanghai Jiao Tong University,
Shanghai 200240 China}

\begin{abstract}
The depletion interactions between two ellipsoids in three configurations were studied
by both Monte Carlo simulation with the Wang-Landau algorithm and the density functional theory in the curvature expansion approximation. Common features of the depletion interactions were found and the results were as expected. By comparing the results 
of the two methods, it is concluded that density functional theory under the curvature expansion approximation gave very good results to the depletion forces.
\end{abstract}
\maketitle

\section{INTRODUCTION}  \label{intro}

Since the work of Asakura and Oosawa(AO)\cite{key-1} who pointed out that
two hard spheres attract each other when suspended in a polymer solution
based on the exclude volume argument over half a century ago, the
researches on the depletion interactions in colloidal suspensions
and in polymer-colloid systems are of longstanding and continuing
interest. As we now understood, AO model is in fact the  first order
approximation in the density of the small component, which was polymer
coils in AO's theory. Mao et al\cite{key-2} extended the AO model to second-
and third-order contributions of the small-particle density, where
 a repulsive potential barrier was found at separations next to the
attractive potential. In subsequent studies, the depletion interactions
of some relatively simple models, induced by a fluid of small hard
spheres, have been studied in detail in theory\cite{key-3, key-4,key-5,key-6}, simulations\cite{key-7, key-8},
and experiments\cite{key-9, key-10}, including interactions between two spheres
or a sphere near a wall. In recent years, complex models were also
studied.  The depletion interactions of big anisotropic objects including
a hard rod or a hard ellipsoid near a planar wall or two
hard spherocylinders have been calculated\cite{key-11,key-12,key-13}. Apart from the
depletion interactions induced by fluids of small spheres, researches
has also been performed to understand the depletion interaction induced
by non-spherical colloidal suspensions. K$\ddot{\textrm{o}}$nig\cite{key-14} studied the depletion
force between non-spherical objects by extending the insertion approach
of Roth\cite{key-6}. Depletion potentials in colloidal mixtures of spheres
and rods have been studied by using the density functional theory (DFT)\cite{key-15}and
Monte Carlo simulations\cite{key-16}. Very recently,  Jin and Wu\cite{key-17}proposed
a hybrid MC-DFT Method for studying depletion interactions,  and used
it to capture the entropic force between asymmetric particles.

The depletion interaction between nonspheric objects and induced
by non-spheric suspensions are important for the real systems where
depletion interaction play a role are mostly consisting of non-spheric
objects. The shape of the ellipsoid may represent a large class of
non-spheric objects range from a needle to a plate by changing the
geometric parameters. Thus in this study we will focus on the depletion
interaction of a simpler problem of two rotational ellipsoids in small
sphere fluid system and try to get a deeper understanding of the 
naeure of depletion.

In this paper, we consider the model  of two hard rotational ellipsoids
in a fluid of small hard spheres. The Monte Carlo simulation with
 Wang-Landau sampling\cite{key-18} is used to calculate the depletion potentials.
Depletion toques are obtained by numerical differentiating the depletion
potential with respect to angles. We also employed the DFT approach
to calculate the depletion potential to compare with our simulation
results.

The relevant details of the models we consider in this work are given
in Sec. II. The implementation of Wang-Landau sampling in our simulation
is presented in Sec. III. In Sec. IV DFT approach under curvature
expansion is described. In Sec. V, we present the simulation results,
and compare with the results obtained by DFT approach. Finally, we
conclude the whole paper in Sec. VI.

\section{Definition of the typical configuration  } \label{define}

The system we shall investigate consists of two rotational ellipsoids
immersed in a hard sphere fluid, one of the ellipsoids is in a fixed position, by
changing the separation and orientation of the second ellipsoid with
the first one and calculating the depletion forces in each configuration,
the full depletion force between the two ellipsoid can be obtained. There
are four degrees of freedom in this  system, the full calculation
amount to scan discretized points in four dimensions, which is a task  too
heavy to be done at the present computation power.  Thus we choose
three typical configurations to study, each of which is specified by a pair of degrees
of freedom, and the three configurations can give a scratch or overall view
of the full depletion force.  The unit of length in  calculation is the diameter of
the small spheres. The ellipsoid is specified by its   long axis $2A$
and short rotational axis $2B$, and in the following the long axis of the fixed ellipsoid is along
the $x$ axis.
The first configuration (referred to as C-1
hereafter)  is  to put the long axis of the second ellipsoid  parallel to the first one, with its center  shifted $x_{||}$ in $x$ direction to the center of the first one, and separated
$h+2B$ in $z$ direction. Thus the two ellipsoids  located at the positions $\left(-\frac{x_{||}}{2},0,-\frac{\left(2B+h\right)}{2}\right)$
and $\left(+\frac{x_{||}}{2},0,+\frac{\left(2B+h\right)}{2}\right)$
respectively. The depletion potential $W\left(x_{||},h\right)$
between these two hard ellipsoids is a function of the shift $x_{||}$ and
the separation $h$ in $z$ direction. The size of the simulation box
is $L_{x}\times L_{y}\times L_{z}$ with $L_{x}=2A+x_{||}+8\sigma$,$L_{y}=2B+8\sigma$
and $L_{z}=4B+h_\text{max}+8\sigma$ respectively. $h_\text{max}$ is the maximum
value of $h$ in our calculation. The choice of simulation box is according
to Dickman\cite{key-8}, which is large enough to neglect finite size effects
safely when volume fraction of the small spheres is less than $0.3$.
The volume fraction of small spheres is defined by $\eta_{s}=N\pi\sigma^{3}/6/V_\text{eff}$
and $V_{\text {eff}}=L_{x}\times L_{y}\times L_{z}-2V_{ellipsoid}$, $V_{ellipsoid}=\frac{4}{3}$$\pi AB^2$ respectively.
Periodic Boundary conditions are applied to all three space directions.

The second configuration(referred to as C-2 hereafter) is that the second ellipsoid
placed $h+2B$ away along the  $z$-axis from the first one, and the  long
axes is rotated about $z$ axis by an angle $\theta$. The two unit vectors
along each long axis are $\left(\sin\left(-\frac{\theta}{2}\right),\cos\left(-\frac{\theta}{2}\right),0\right)$
and $\left(\sin\left(+\frac{\theta}{2}\right),\cos\left(+\frac{\theta}{2}\right),0\right)$
respectively. The centers of mass of the two hard ellipsoids are located
at $\left(0,0,\frac{-\left(2B+h\right)}{2}\right)$ and $\left(0,0,\frac{\left(2B+h\right)}{2}\right)$
respectively. The size of the simulation box is $L_{x}\times L_{y}\times L_{z}$
with $L_{x}=2B+2A\sin\left(\frac{\theta}{2}\right)+8\sigma$,$L_{y}=2B+2A\cos\left(\frac{\theta}{2}\right)+8\sigma$
and $L_{z}=4B+h_\text{max}+8\sigma$ respectively, chosen with the same criterion
as that in C-1. The depletion potential $W\left(\theta,h\right)$
depends on the angle $\theta$ and the separation $h$ of the two hard
ellipsoids. The periodic boundary conditions are also applied to all
three space directions.

The third configuration (referred to  as C-3 hereafter) is specified by two
parameters, the angle $\Phi$ of the second  ellipsoid rotated about $y$ axis
and minimal surface-to-surface distance $h$ between two hard ellipsoids.
For example, $h=0$ when the two hard ellipsoids just contact each other.
$z_\text{min}$ and $x_\text{min}$ depend on angle $\Phi$ by $z_\text{min}=\left(A^{2}\sin^{2}\left(\frac{\Phi}{2}\right)+B^{2}\cos^{2}\left(\frac{\Phi}{2}\right)\right)^{\frac{1}{2}}$
and $x_\text{min}=\left(A^{2}\cos^{2}\left(\frac{\Phi}{2}\right)+B^{2}\sin^{2}\left(\frac{\Phi}{2}\right)\right)^{\frac{1}{2}}$
from simple geometric considerations. When $h$ and $\Phi$ are given, the
locations of the two hard ellipsoids are at $\left(0,0,\frac{-\left(2z_\text{min}+h\right)}{2}\right)$
and $\left(0,0,\frac{\left(2z_\text{min}+h\right)}{2}\right)$.  The unit
vectors along each hard ellipsoid are $\left(\cos\left(\frac{\Phi}{2}\right),0,\sin\left(\frac{\Phi}{2}\right)\right)$
and $\left(\cos\left(-\frac{\Phi}{2}\right),0,\sin\left(-\frac{\Phi}{2}\right)\right)$.
According to the same criterion as the first two configurations, the simulation
box size is chosen to be $L_{x}*L_{y}*L_{z}$with $L_{x}=2x_\text{min}+8\sigma$, $L_{y}=2B+8\sigma$,$L_{z}=4*z_\text{min}+h_\text{max}+8\sigma$.The periodic
boundary conditions are used in three space directions.

\section{Method of calculation} \label{method}
\subsection{Monte Carlo Simulation}  \label{MC}

We describe the simulation method here. The depletion potential for a given configuration
of the two ellipsoids is the free energy of the whole system under such a configuration, we
may chose the zero point of the depletion potential when the two ellipsoids are infinitely separated.
Thus what we need to calculate is the free energy difference of a configuration with that of infinite separation.
There are different  schemes  in literatures in the evaluation of the free energy difference by
Monte Carlo simulations. Here we adopt the Wang-Landau  method.
The method was first proposed by Wang and
Landau\cite{key-18} with lattice models to calculate the energy density of states, and extended to off-lattice systems by Shell\cite{key-19}. The efficiency  and robustness  of the Wang-Landau sampling algorithm in estimation
of the density of state(DOS) are excellent in a wide range of model systems.

The system in this study consists of hard objects only, thus the internal energy is only the
kinetic energy which is determined by temperature only and irrelevant to the depletion potential.
The part that contributes to depletion potential is the variance of entropy of the system with
configurations and the entropy is directly related to the number of microscopic states as
$S=k_B \ln g$ where $g$ is the number of microscopic states of a given macroscopic configuration.
The depletion potential $W$ is given by
\[
\frac{W}{k_B T} =-\left(\frac{S}{k_B} -\frac{S_0}{k_B}\right) =-\ln\left(\frac{g}{g_0}\right).
\]
Here $S_0$ and $g_0$ are the entropy and number of microscopic states of the system when two
ellipsoids infinitely separated, respectively.
The ratio of the number of microscopic states can be calculated directly from the Wang-Landau method.
In the three macroscopic configurations considered, each configuration was characterized by two parameters.
The calculation is proceeded in the following way: one parameter is fixed at several different values and the second parameter is divided into small intervals, the microscopic density of states of the second parameter is then
evaluated by the Wang-Landau method and the depletion potential is then calculated.  For each fixed value of
the first parameter, an independent run is needed.

The sampling of the simulation consists of two parts. The first part is the move between ellipsoids configurations,  the second ellipsoid moved in the one parameter intervals
according to the acceptance
criterion $P(State_{old}\rightarrow State_{new})=\text{min}(1,g(State_{old})/g(State_{new}))$. where
$g(State)$ is the density of states of the current state, specified by the fixed first parameter and the current interval of the second parameter.   When a state is visited, the corresponding density of states $g(state)$ is updated by
multiply a factor $f$. The second part is the sampling of the small hard spheres when an ellipsoids configuration is specified,  the Metropolis sampling is used in this part, i.e. a randomly chosen small hard sphere with a random trial move is accepted if the move does not result an overlap with other objects. For each move of the ellipsoid, usually $10^{5}$ Monte Carlo steps of the small hard sphere system were performed in order to keep the equilibrium.  The initial value of $f=f_{0}$  is  chosen to be $e^{0.01}$ in our simulation, which is different with
the common value $e^{1}$,  simply because the free energy landscape of depletion is relatively
flat compared with other systems. Accumulated histogram $H(state)$ are
updated during the random walk. When $H$ is flat, we clear $H$ and modify
$f$ to $f^{1/2}$.  The whole simulation stopped  when the modification
factor smaller than $exp(10^{-8})$.

\subsection{DENSITY FUNCTIONAL THEORY}  \label{DFT}

In this section we describe the density functional theory (DFT) method used in this study.
The DFT is a very efficient theoretical method for
the calculation of depletion interactions.  The basics of the method is to take the two ellipsoid as
external potential of the small hard sphere systems and finding  the density distribution and free
energy of the small hard sphere system by the minimization of the free energy functional. For hard sphere systems
the fundamental measure theory(FMT) proposed long time ago by Rosenfeld  and its various extensions and modifications usually gave pretty accurate values of free energy. However, the minimization of the free energy
functional in the general three dimension is computationally heavy and, in some cases even impractical. In
real calculations one often reduce the problem to two dimensional or even one dimensional by symmetry of the
system studied. In the depletion potential calculation, the so called
insertion approach\cite{key-6} proposed by R. Roth et al is  a very effective way to reduce the dimensionality. In this approach only  one solute particle is fixed so that higher symmetry usually available and  thus it can drastically reduce the computation load.

In the insertion approach the depletion potential is related to the one body direct correlation function $C_{b}^{\left(1\right)}\left(\mathbf r, \omega\right)$  as
\begin{equation}
\beta W\left(\mathbf r,\omega\right)=C_{b}^{\left(1\right)}\left(|\mathbf r | \to  \infty,\omega\right)-C_{b}^{\left(1\right)}\left(\mathbf r,\omega\right),
\end{equation}
 the direct correlation function can be calculated within FMT framework by \cite{key-20}
\begin{equation}
C_{b}^{\left(1\right)}=-\sum_{\alpha}\frac{\partial\Phi}{\partial n_{\alpha}}\bigotimes\omega_{\alpha}^{b}.
\end{equation}
Here  $\Phi$ is a  function of the weighted density  $\left\{ n_{\alpha}\left(r\right)\right\} $,
 given by
\begin{equation}
\Phi\left(\left\{ n_{\alpha}\left(\mathbf r\right)\right\} \right)=-n_{0}ln\left(1-n_{3}\right)+\frac{n_{1}n_{2}-\mathbf{n_{1}\cdot n_{2}}}{1-n_{3}}+\frac{n_{2}^{3}-3n_{2}\mathbf{n_{2}\cdot n_{2}}}{24\pi\left(1-n_{3}\right)^{2}}
\end{equation}
where $\left\{ n_{\alpha}\right\} $ are  weighted densities of  four scalar types and two
vector types given by:
\begin{equation}
n_{\alpha}\left(\mathbf r\right)=\int\omega_{\alpha}\left(\mathbf r-\mathbf r^{'}\right)\rho\left(\mathbf r^{'}\right)d\mathbf r^{'},\quad \alpha=0,1,2,3
\end{equation}
 \begin{equation}
\mathbf{n_{\alpha}}\left(r\right)=\int\bm{\omega_{\alpha}}\left(\mathbf r-\mathbf r^{'}\right)\rho\left(\mathbf r^{'}\right)d\mathbf r^{'}, \quad \alpha=1,2.
\end{equation}
The weight functions for spherical hard body are given by
 
\begin{equation}
w_{3}\left(\mathbf{r}\right)=\text{$\Theta$}\left(|\mathbf{r}-\mathbf{R}_{b}\left(\theta,\phi\right)|\right)
\end{equation}

\begin{equation}
w_{2}\left(\mathbf{r}\right)=\delta\left(|\mathbf{r}-\mathbf{R}_{b}\left(\theta\text{,}\phi\right)|\right)
\end{equation}

\begin{equation}
\omega_{1}\left(\mathbf{r}\right)=\frac{\omega_{2}}{4\pi R}
\end{equation}
 \begin{equation}
\omega_{0}\left(\mathbf{r}\right)=\frac{\omega_{2}^{b}}{4\pi R^{2}}
\end{equation}
\begin{equation}
{\mathbf w_{2}\left(\mathbf{r}\right)}=-\nabla\omega_{3}\left(\mathbf{r}\right)
=\mathbf{\mathbf{n}_{\mathbf{b}}}\left(\mathbf{r}\right)\delta\left(|\mathbf{r}-\mathbf{R}_{b}\left(\theta,\phi\right)|\right)
\end{equation}
\begin{equation}
{\mathbf w_{1}\left(\mathbf{r}\right)}=\frac{ { \mathbf w_{2}\left(\mathbf{r}\right)}}{4\pi R^{2}}
\end{equation}

In our system, under the insert particle scheme, the fixed solute is a rotational ellipsoid, a full calculation  requires the minimization of the free
energy functional on a two dimensional grid, which is still computationally heavy if the grid is refined to reach the
required accuracy.  If it can be further reduced to a one dimensional problem, then the calculation will be much easier to accomplish. In a recent work\cite{key-21}, K$\ddot{\textrm{o}}$nig et al. presented an ansatz for the
density profile of small particles around a big fixed object:
\begin{equation}
\rho_{s}\left(\mathbf{r}\right)=\rho_{s}^{P}\left(u\right)+H\left(\mathbf{R}\right)\rho_{s}^{H}\left(u\right)+K\left(\mathbf{R}\right)\rho_{s}^{K}\left(u\right)+\cdot\cdot\cdot
\end{equation}
$ \mathbf{r}$ is the point outside the fixed object, $\mathbf{R}$ is the
closest point from $\mathbf{r}$ on the surface of the fixed object where $\rho_{s}\left(\mathbf{r}\right)$
vanishes.   $\mathbf{R}-\mathbf{r}=u\mathbf{n(R)}$,
$\mathbf{n(R)}$ is the unit vector normal to the surface at point $\mathbf{R}$.
${H}$ and ${K}$ are   the mean and Gaussian curvature at point
$\mathbf{R}$.
Based on this assumption, K$\ddot{\textrm{o}}$nig et al. then introduced the curvature expansion approximation
for studying depletion force between two nonspherical objects\cite{key-14}.
They further argue that $\Psi_{\alpha}\left(\mathbf{r}\right)$ for
one fixed object can be expanded by the surface curvature of the object
\begin{equation}
\Psi_{\alpha}\left(\mathbf{r}\right)=\Psi_{\alpha}^{P}\left(u\right)+H\left(\mathbf{R}\right)\Psi_{\alpha}^{H}\left(u\right)+K\left(\mathbf{R}\right)\Psi_{\alpha}^{K}\left(u\right)+\cdot\cdot\cdot
\end{equation}
where $\frac{\partial\Phi}{\partial n_{\alpha}}=\Psi_{\alpha}$.  Under
this approximation, we only need to calculate $\Psi_{\alpha}\left(\mathbf{r}\right)$
near simple geometry objects to get $\Psi_{\alpha}^{P}\left(u\right)$,  $\Psi_{\alpha}^{H}\left(u\right)$, and
$\Psi_{\alpha}^{K}\left(u\right)$. And then $\Psi_{\alpha}\left(\mathbf{r}\right)$
near nonspherical object can be obtained.

In order to obtain the convolution of $c_{b}^{\left(1\right)}$ for
the insertion of second nonspherical object, Rosenfeld$^{'}$s generalized
FMT for convex hard bodies\cite{key-22} is employed as Roth$^{'}$s previous
application in the calculation of the depletion torque\cite{key-11}. The weight
functions are
\begin{equation}
w_{3}^{b}\left(\mathbf{r}\right)=\text{$\Theta$}\left(|\mathbf{r}-\mathbf{R}_{b}\left(\theta,\phi\right)|\right)
\end{equation}

\begin{equation}
w_{2}^{b}\left(\mathbf{r}\right)=\delta\left(|\mathbf{r}-\mathbf{R}_{b}\left(\theta\text{,}\phi\right)|\right)
\end{equation}
\begin{equation}
\omega_{1}^{b}\left(\mathbf{r}\right)=\frac{H\left(\mathbf{r}\right)\omega_{2}^{b}}{4\pi}
\end{equation}
\begin{equation}
\omega_{1}^{b}\left(\mathbf{r}\right)=\frac{H\left(\mathbf{r}\right)\omega_{2}^{b}}{4\pi}
\end{equation}
where $H(r)$ is the mean curvature of the second object, and
\begin{equation}
\omega_{0}^{b}\left(\mathbf{r}\right)=\frac{K\left(\mathbf{r}\right)\omega_{2}^{b}}{4\pi}
\end{equation}
where $K(\mathbf r)$ is the Gaussian curvature of the second object.
\begin{equation}
{\mathbf w_{2}^{b}\left(\mathbf{r}\right)}
=-\nabla\omega_{3}^{b}\left(\mathbf{r}\right)
=\mathbf{\mathbf{n}_{\mathbf{b}}}\left(\mathbf{r}\right)\delta\left(|\mathbf{r}-\mathbf{R}_{b}\left(\theta\text{,}\phi\right)|\right)
\end{equation}
where $\mathbf{\mathbf{n}_{\mathbf{b}}}\left(\mathbf{r}\right)$ is
the unit vector normal to the surface at point  $\mathbf{r}$,
and
\begin{equation}
 {\mathbf w_{1}^{b}\left(\mathbf{r}\right)}
=\frac{{H\left(\mathbf{r}\right) \mathbf w_{2}^{b}\left(\mathbf{r}\right)}}{4\pi}
\end{equation}

\section{RESULTS AND COMPARATION}
We carried out calculations both with Monte Carlo simulations and DFT method described in the previous section for the three configurations defined in section \ref{define}.

\begin{figure}
\begin{minipage}{0.5\textwidth}
\includegraphics[width=0.9\textwidth]{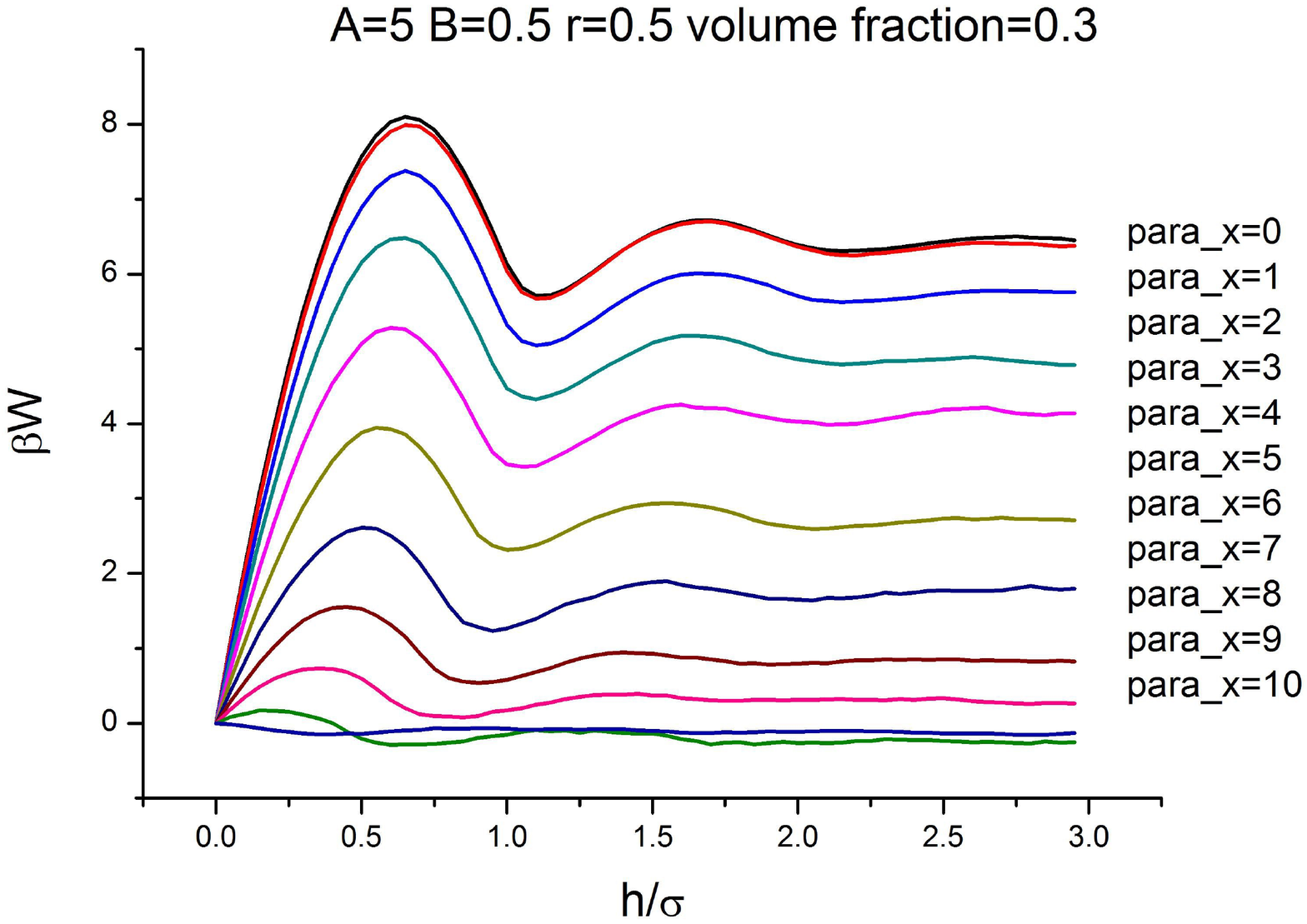}   \\
\centering{(a)}
\end{minipage}\hfil \begin{minipage}{0.5\textwidth}
\includegraphics[width=0.9\textwidth]{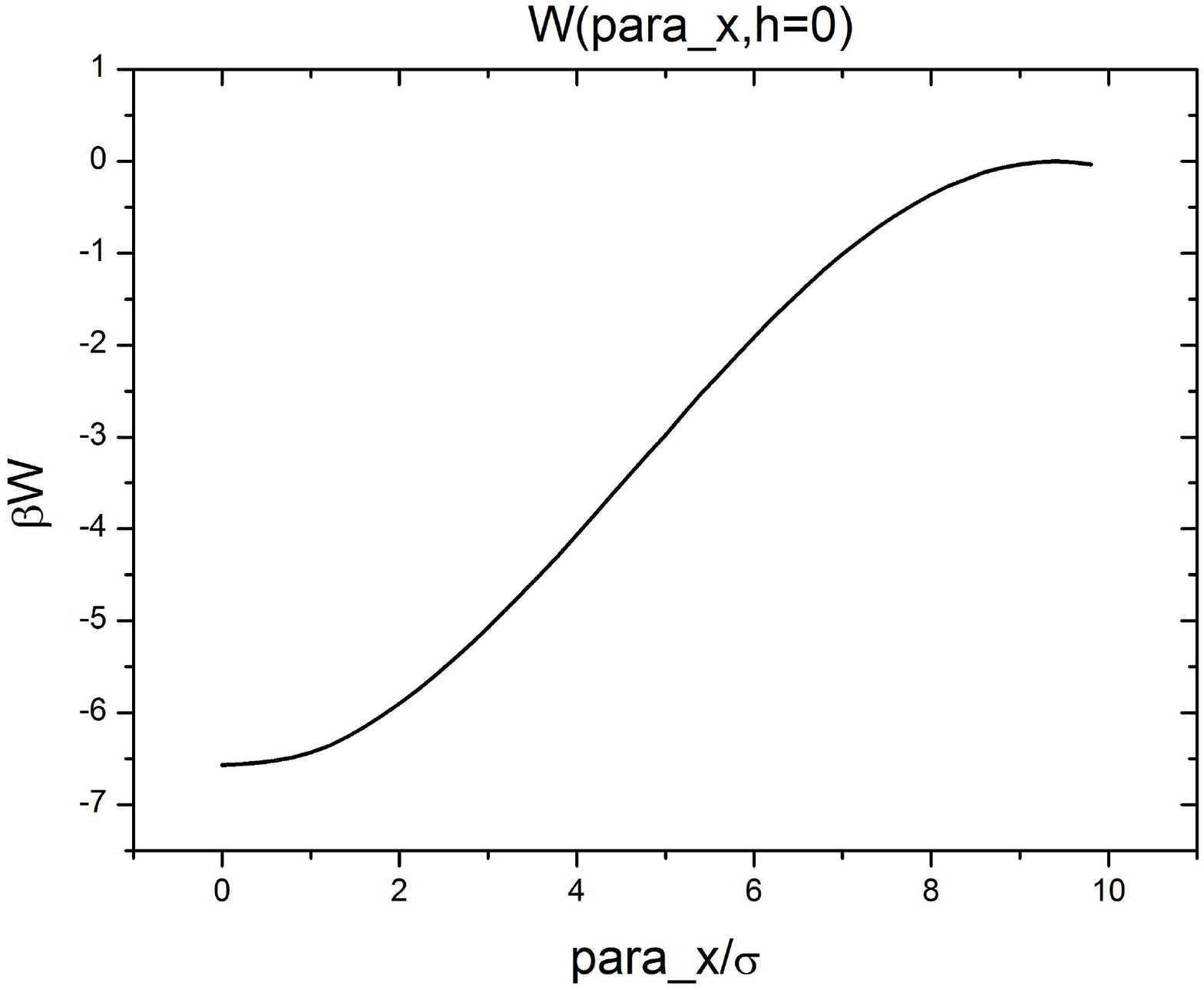}  \\
\centering{(b)} \\
\end{minipage}
\caption{(a), The depletion potential of configuration C-1, $A=5\sigma$, $B=0.5\sigma$, the volume fraction of small hard spheres is $\eta_s = 0.3$. Each curve corresponding to one $x_\parallel$ and all curves are shifted so that the $W(x_\parallel, 0) =0$ for clarity. (b), The variation of depletion potential $W(x_\parallel, 0)$ with respect to $x_\parallel$. } \label{fig1}
\end{figure}

Figure \ref{fig1} (a) shows the Monto Carlo simulation results of the variation of depletion potential $\beta W\left(x_{||},h\right)$
with $h$ for different fixed  $x_\parallel$ of the configuration C=1. The parameters for the ellipsoids are  $A=5\sigma$ and $B=0.5\sigma$. The
volume fraction of small hard spheres is $\eta_{s}=0.3$. The curves are the variations of the depletion potential with the separation between  the ellipsoids,   each curve corresponding to a  fixed value of $x_\parallel$. For clarity reasons,    all the potential  curves are shifted so that their potential zero at the point $\left(x_{||},0\right)$.  From the figure we see that the depletion force is much stronger when the centers of the two ellipsoids  are not shifted with each other, and becomes weaker when the shift becomes larger. The
properties of the depletion force is similar with different shifts and also similar to the depletion potential
between two hard spheres, i.e, the depletion force is attractive at small separations and then turn to repulsive at
a separation about the diameter of small hard spheres, and oscillates slightly afterwards then tends to zero.
This is a typical feature of depletion.
 Figure \ref{fig1} (b) is the variation of the depletion potential with the shift $x_\parallel$ at $h=0$, it is precisely the
relative shift value of each curve in figure \ref{fig1} (a) .   The curve is relative flat in the regions $x_{||}<1$
and $x_{||}>8$. In the first case the two ellipsoid are only shifted slightly and the influence of the shift is small, and in the second case the shift is so large that the two ellipsoids are already well separated in the $x$ direction thus further shift only has very small effect on the depletion potential.   In the middle part the shift
changes the depletion domain(the region of space that the small hard spheres are unable to enter) between the two ellipsoids thus changes the depletion potential.

\begin{figure}
\begin{minipage}{0.5\textwidth}
\includegraphics[width=0.9\textwidth]{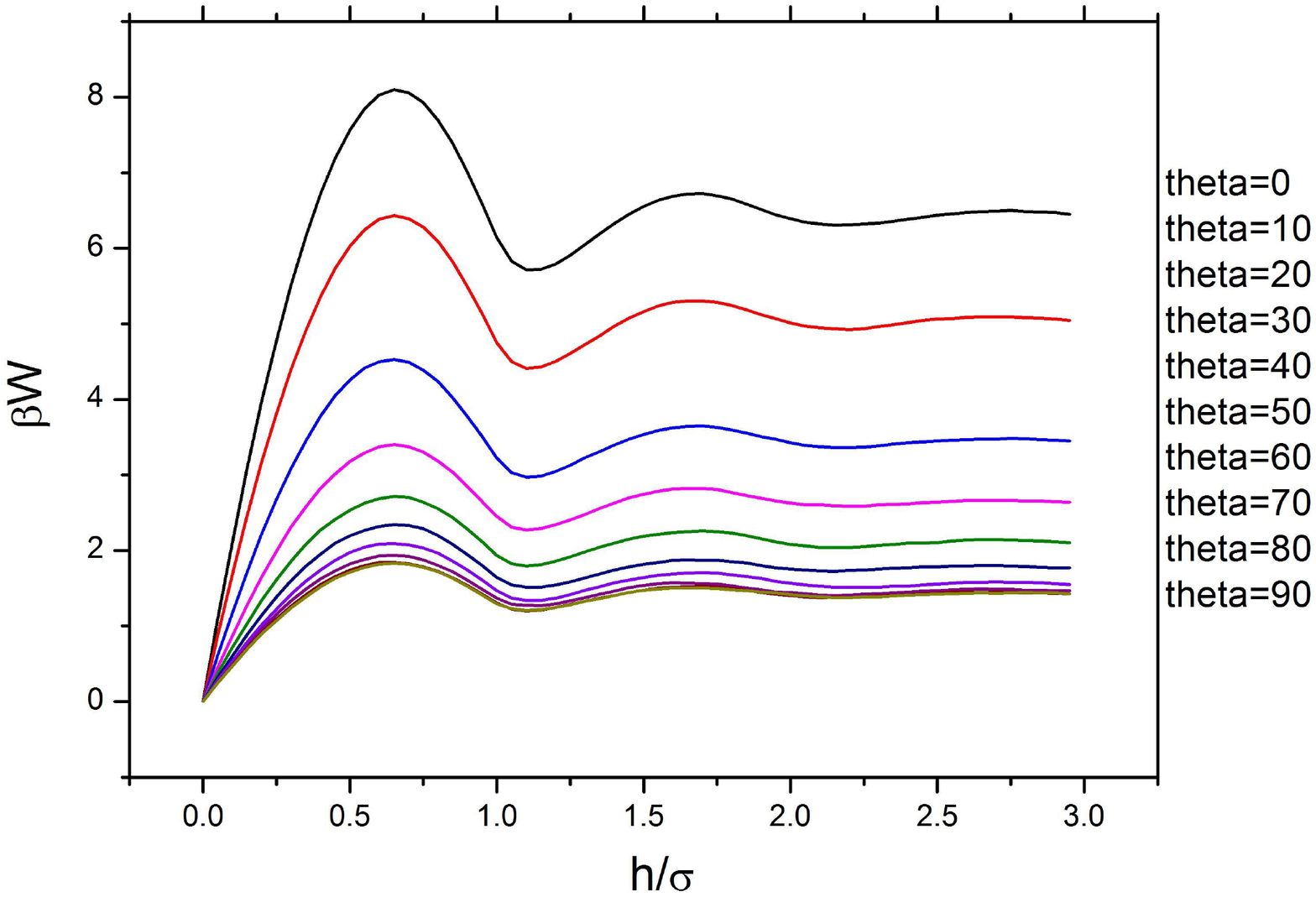} \\
\centering{(a)} 
\end{minipage}\hfil \begin{minipage}{0.5\textwidth}
\includegraphics[width=0.9\textwidth]{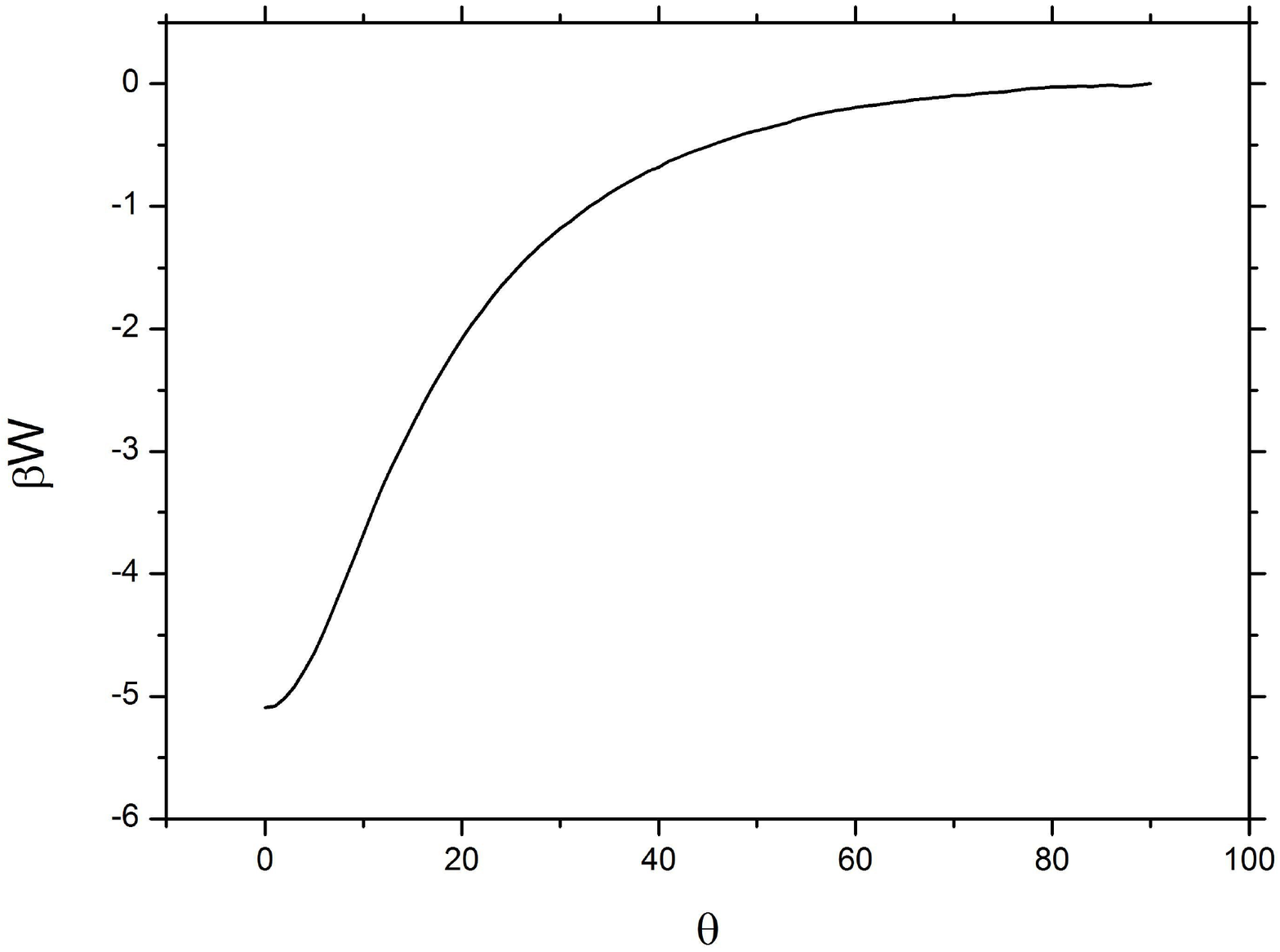}  \\
\centering{(b)}
\end{minipage}

\includegraphics[width=0.45\textwidth]{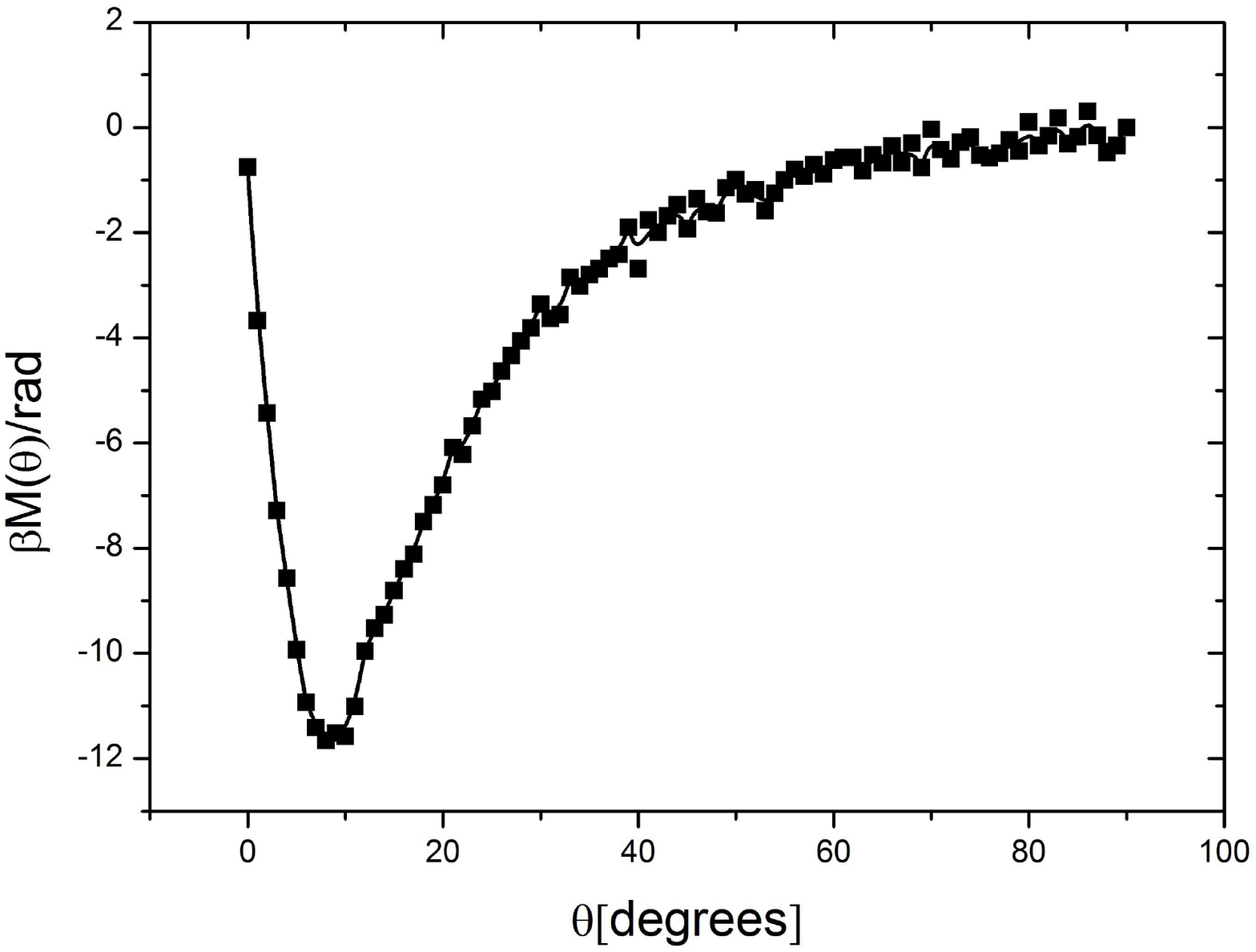}\\
\centering{(c)}\\
\caption{(a) The depletion potential $W(\theta, h)$ of configuration C-2. Each curve corresponding to a fixed value of rotation angle $\theta$, the curves are shifted so that the zero of the potential is at point $h=0$, other parameters are the same as the figure \ref{fig1}. (b)  The variation of the depletion potential at $h=0$ with the rotation angle $\theta$. (c) The depletion torque between the two ellipsoids obtained from numerical differentiation of the depletion potential for $h=0$. }  \label{fig2}
\end{figure}

Figure
\ref{fig2} (a) are  the variation with the separation $h$ of the depletion potential $\beta W\left(\theta,h\right)$ for different fixed rotation values of $\theta$. The other parameters are the same as the figure \ref{fig1}. For clarity reasons the curves are  shifted so that the potential at $h=0$ coincide and set to zero for different $\theta$'s.  It is clear that the depletion force in the cases of small $\theta$ is much stronger than those of large $\theta$, which is the reflection of the fact that for the long ellipsoids studied here, small $\theta$ means strong depletion and results larger depletion force.   The variation of depletion potential $W\left(\theta,h=0\right)$ with rotation angle $\theta$ at $h=0$ is given in figure \ref{fig2} (b).  It variates strongly in the small $\theta$ region and tends to be flat for large $\theta$ as expected.  Figure \ref{fig2} (c)
is the depletion torque at $h=0$ obtained from numerical differentiation of the depletion potential as shown in
figure \ref{fig2} (b).  The torque is less than zero which means it is a restoring torque as already clear from the
depletion potential curve of figure \ref{fig2} (b). This restoring torque is at its maximum at $\theta\approx 8^\circ$.

 \begin{figure}
\begin{minipage}{0.5\textwidth}
\includegraphics[width=0.9\textwidth]{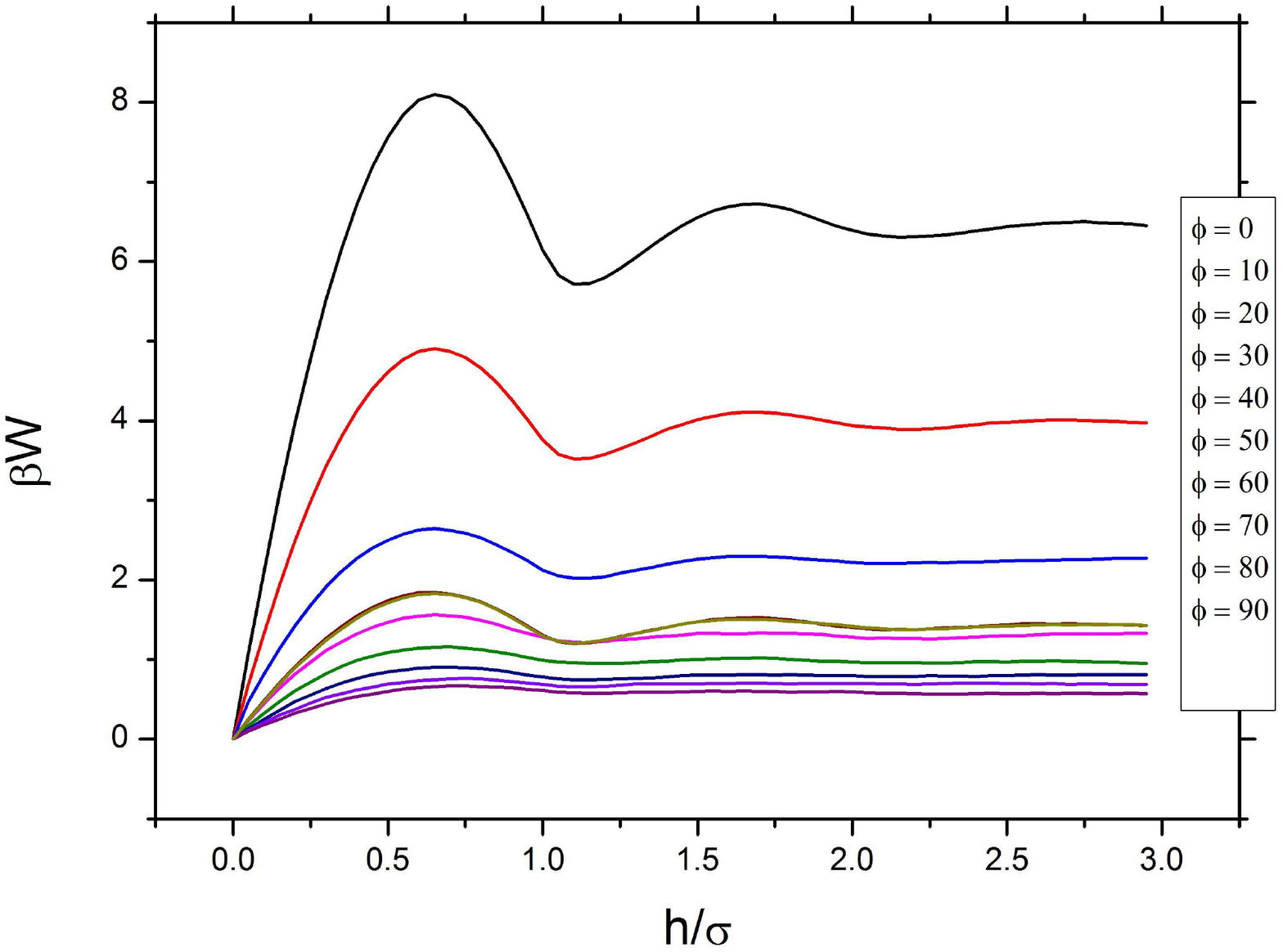} \\
 \centering{(a)} \\ 
\end{minipage}\hfil \begin{minipage}{0.5\textwidth} 
\includegraphics[width=0.9\textwidth]{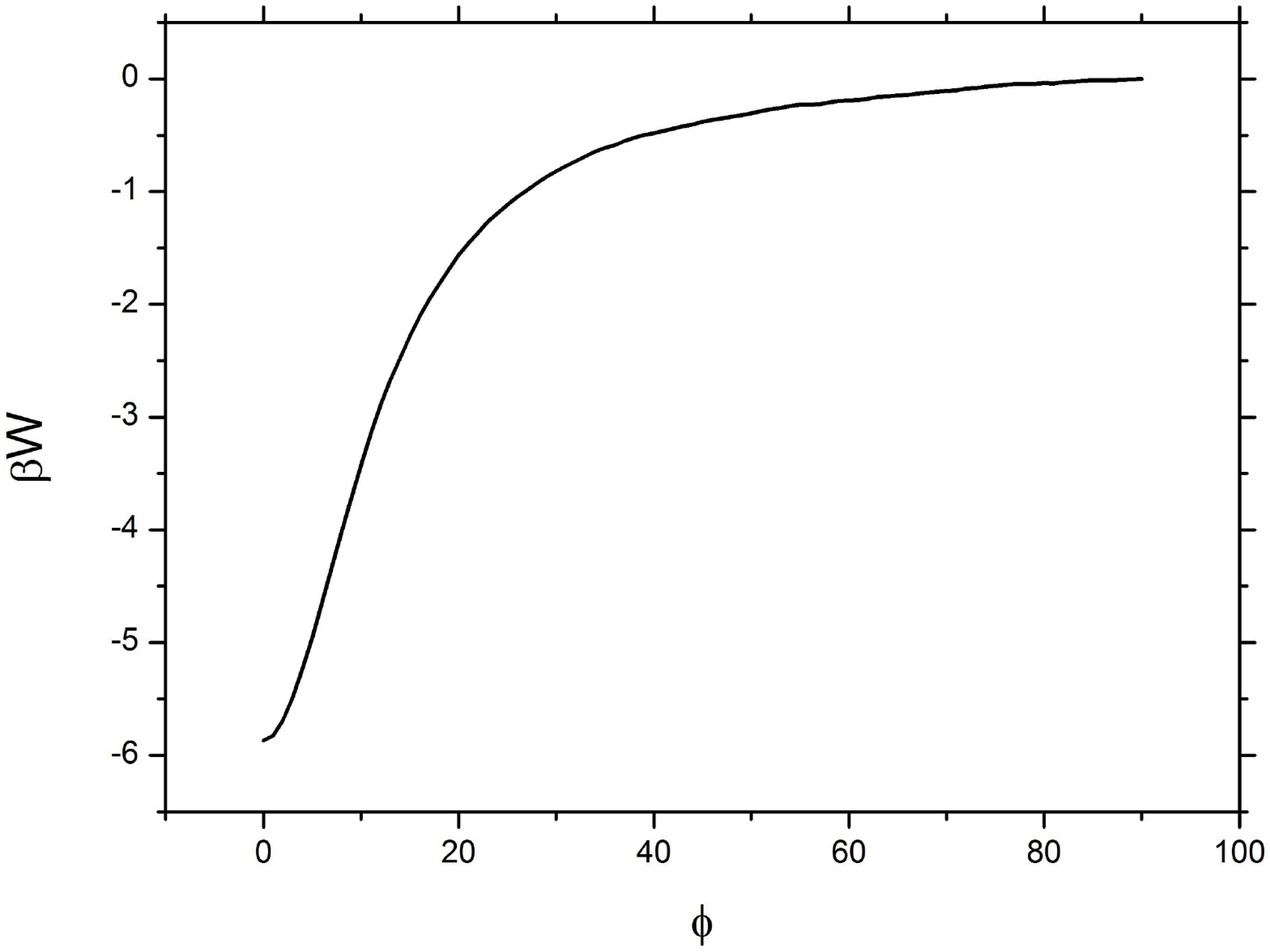}  \\
\centering{(b)}  \\
\end{minipage}
\includegraphics[width=0.45\textwidth]{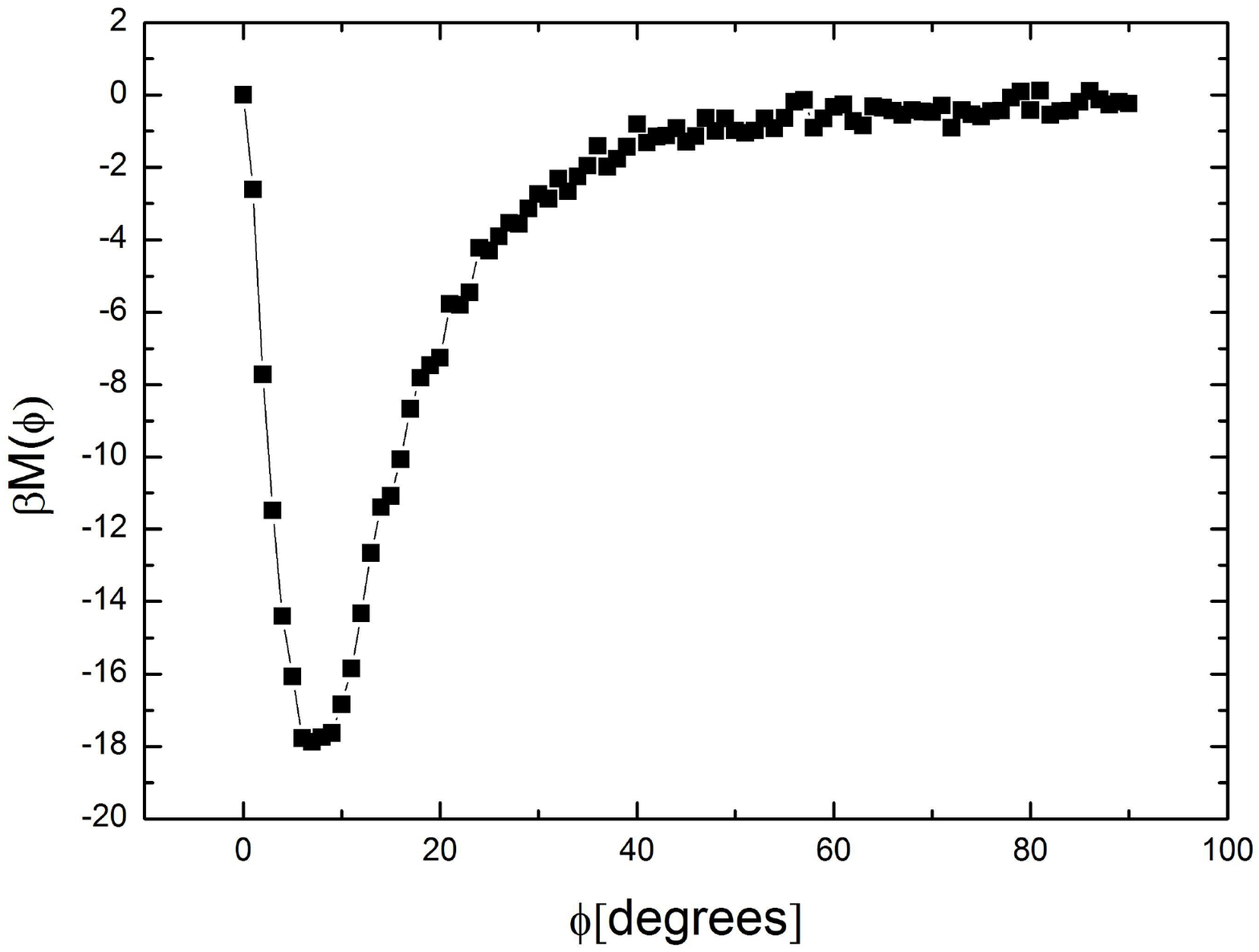}  \\
\centering{(c)} 
\caption{(a) The variation of  depletion potential with $h$ for configuration 3 with different fixed values of $\Phi$, the curves are shifted in the same way as figure \ref{fig1}  and the parameters are also the same as figure \ref{fig1}.  (b),  The variation of  depletion potential with $\Phi$ at $h=0$. (c) The depletion torque at
$h=0$ as function of $\Phi$. }\label{fig3}
\end{figure}

The depletion potential of the configuration C-3,  $\beta W\left(\Phi,h\right)$
is shown in figure \ref{fig3}. Figure \ref{fig3} (a) shows the variation of depletion potential with $h$ for
several fixed $\Phi$'s. The other parameters  are the same as
in figure \ref{fig1}. The curves are shifted in the same way as figure \ref{fig1} and \ref{fig2} so that the
potential zero is at $h=0$. The variation of the depletion potential with $\Phi$ at $h=0$ is shown in figure
\ref{fig2} (b) and the depletion torque with respect to the $y$ axis, obtained from numerical differentiation of the figure {\ref{fig3} (b),  is shown in figure \ref{fig3} (c).  As in the case of configuration C-2, the torque is negative and has the effect to restoring the configuration to the $\Phi =0$ state.

Put  together the results of all three configurations we can conclude that the minimum of the
depletion potential is the state  that two ellipsoids parallel in their long axis without shifting.
When the two ellipsoids deviates from such a state the depletion force and depletion torque will
restore the state. Since the three body depletion force and high order many body depletion forces are  very
small compared to two body ones, it is expected that the depletion   may play  a crucial role for the self-assembly  of ellipsoids system  to form nematic order when immersed in a  small hard sphere system or in a solution with polymer blends.
 
\begin{figure}
\begin{minipage}{0.5\textwidth}
\includegraphics[width=0.9\textwidth]{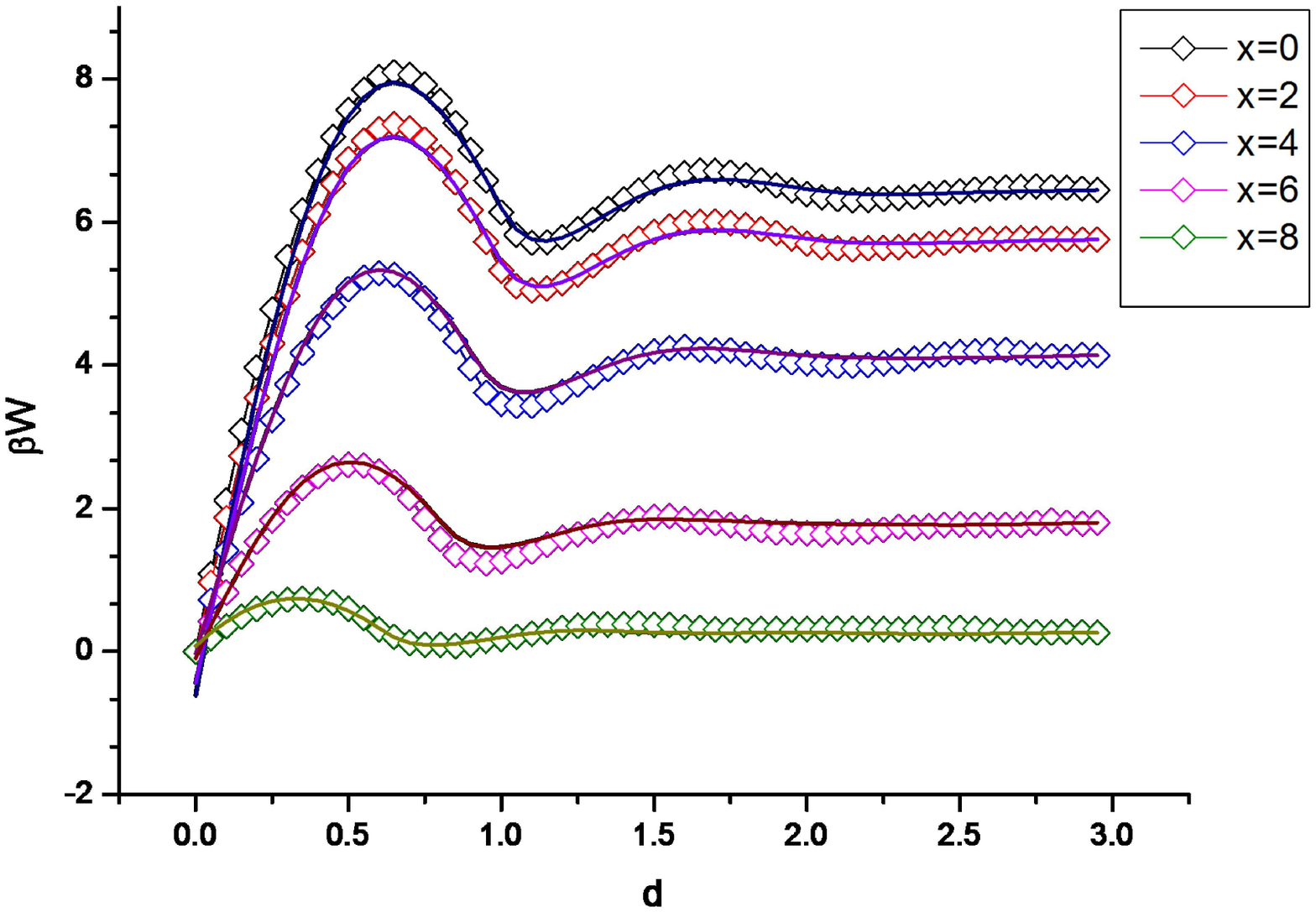} \\
 \centering{(a)} \\ 
\end{minipage}\hfil \begin{minipage}{0.5\textwidth} 
\includegraphics[width=0.9\textwidth]{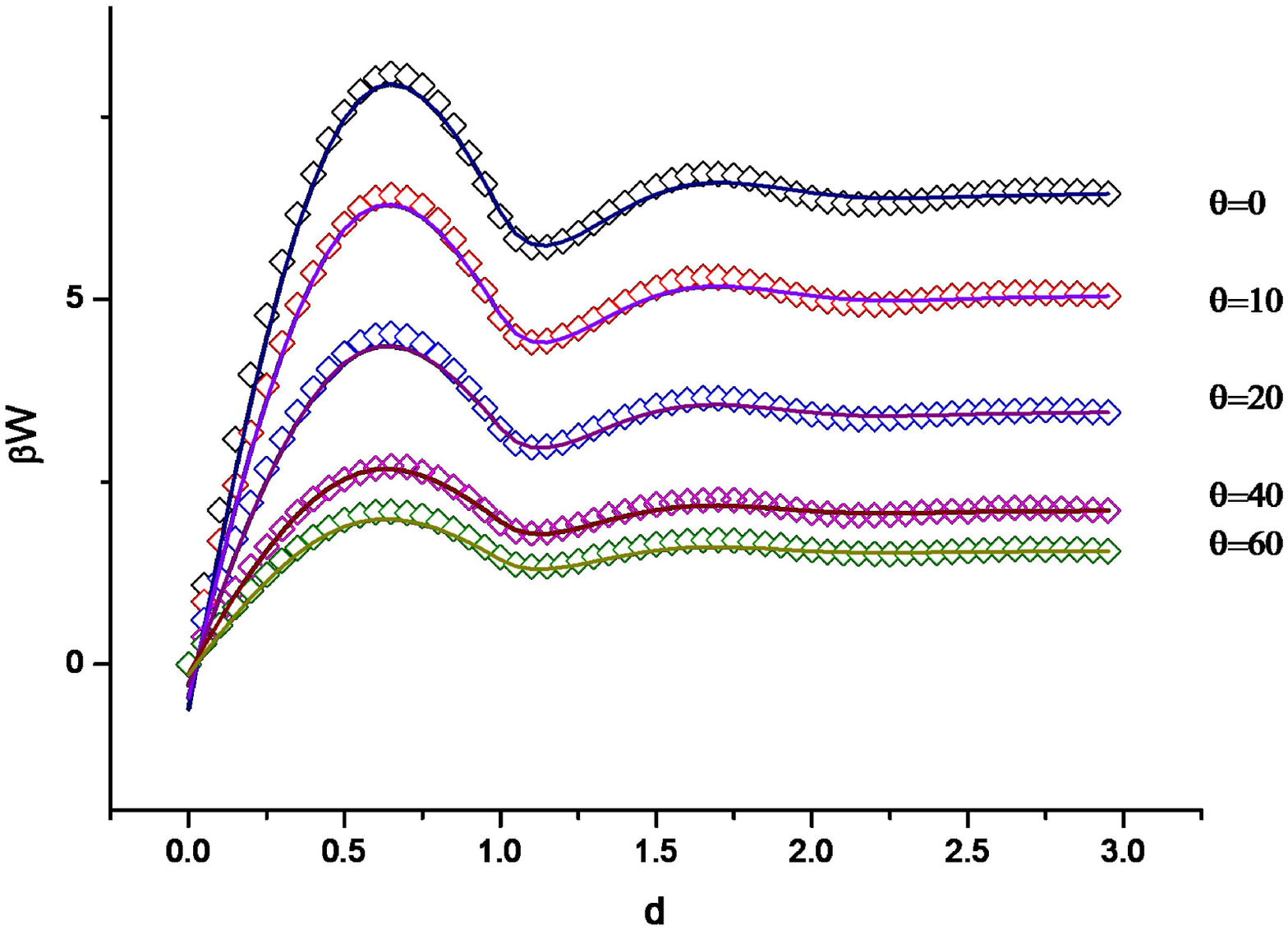}  \\
\centering{(b)}  \\
\end{minipage}
\includegraphics[width=0.45\textwidth]{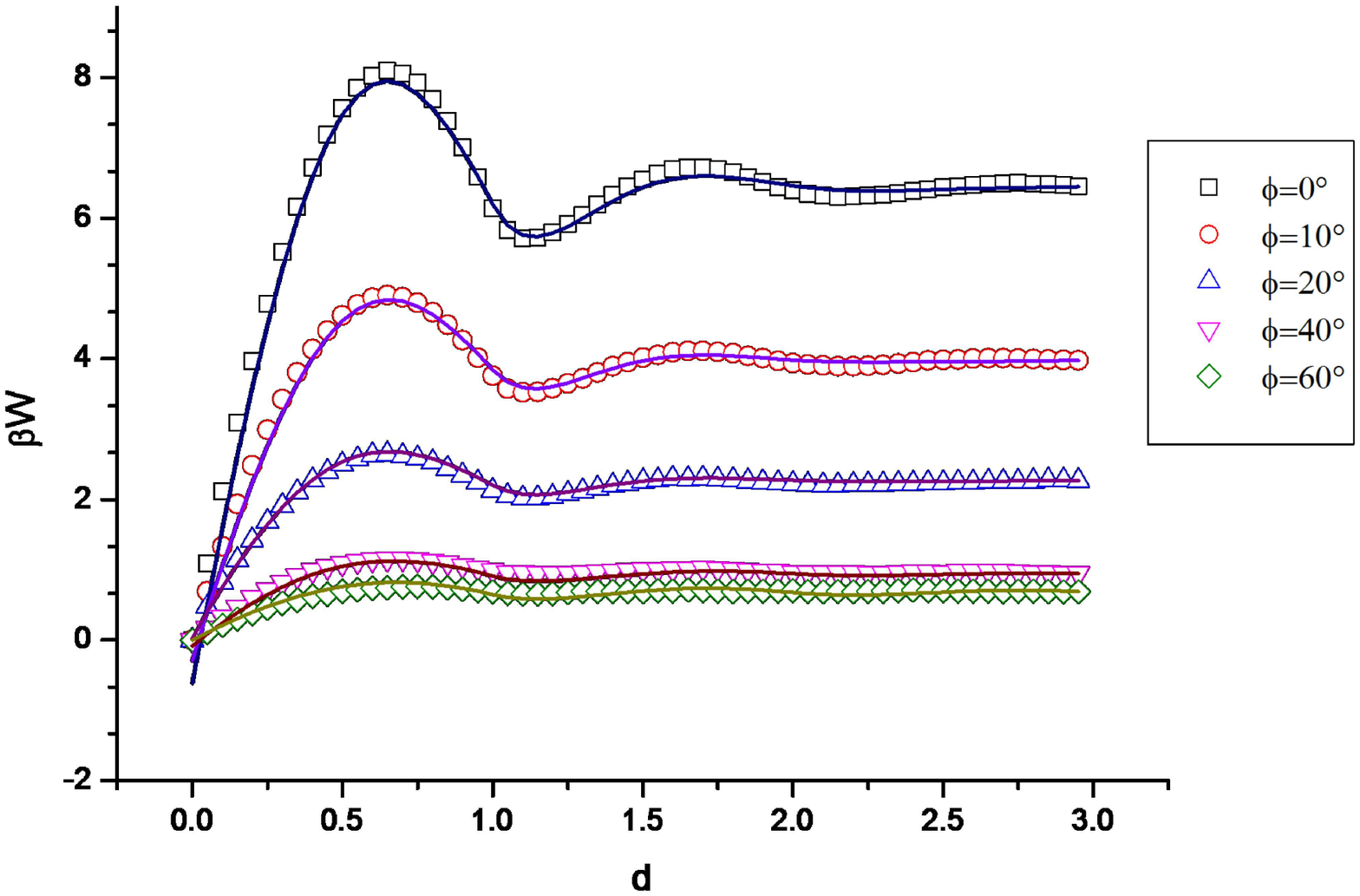}  \\
\centering{(c)} 
\caption{Comparison of the depletion potentials calculated from MC simulations(symbols) and  from DFT curvature expansion method(solid lines). } \label{fig4}
\end{figure}

Figure \ref{fig4}(a)  --- figure \ref{fig4}(c)  are the comparisons of the DFT results under the
curvature expansion approximation with the Monto Carlo results.  Figure \ref{fig4} is for the
configuration C-1, where symbols are results from MC simulations and lines are from DFT method
in the curvature expansion as described in section \ref{DFT}.

 Figure \ref{fig4}(a)  and Figure \ref{fig4}(b) show the comparisons of MC and DFT  for configuration C-2 and configuration C-3  respectively. The agreement between DFT results
and the corresponding simulation results are considerably good.  As indicated by K$\ddot{\textrm{o}}$nig\cite{key-14,key-21},  the most inaccurate part of the curvature expansion
approximation is  those region near the surface of the fixed object.
Thus our model of two nearly parallel needle-like ellipsoids near each
other is a very harsh test situation for the approximation, since the
near surface contribution is the main contribution  of the whole integral. Even
though, inaccuracy of the depletion well depth obtained by the DET under curvature
expansion approximation is less than 10\%.  Since the curvature expansion DFT method
use much smaller computational resources compared to both the original FMT DFT method
and the Monte Carlo simulation, and it is also easy to implement, it is an excellent method
in the estimation of depletion forces for many different situations of non spherical objects.

\section{SUMMARY}

In summary, we calculated the depletion interaction of two hard ellipsoids
in a fluid of small hard spheres by using Wang-Landau sampling Monte
Carlo simulations and the density functional theory under curvature expansion.
Our simulation suggests that Wang-Landau sampling is an efficient method
for calculating the depletion potential of two ellipsoids, and the
curvature expansion DFT approach is a much computationally cheaper yet
considerably accurate theoretical method compared with the simulation
ones.  Instead of an investigation of the depletion potential in the whole parameter space, which is
both unnecessary and impractical for its heavy computation load,    we have revealed
the key aspects of depletion interactions in this system by choosing
the three representative configurations. 

Work supported by the National Nature Science Foundation of China under grant \#10874111,  \#11304169 and
\#11174196.

\end{document}